\documentstyle[amssymb,preprint,aps]{revtex}
\tightenlines

\begin{document}
\title{The Equivalence Principle and gravitational and inertial mass relation of
classical charged particle}
\author{Mario Goto}
\address{Departamento de F\'{\i }sica/CCE\\
Universidade Estadual de Londrina\\
86051-990 Londrina, PR - Brazil\\
(mgoto@uel.br)}
\maketitle
\date{}

\begin{abstract}
We show that the locally constant force necessary to get a stable hyperbolic
motion regime for classical charged particles, actually, is a subtle
combination of an applied external force and the radiation reaction force.
It suggests, as the Equivalence Principle is valid, that the gravitational
mass of charged particle should be slight greater than its inertial mass.
However, an interesting new feature emerges from the unexpected behavior of
the gravitational and inertial mass relation at very strong gravitational
field, just reachable at the early stage of the universe or, perhaps, near
massive black holes, with a divergence that introduces a critical field
value. It signs that the Equivalence Principle should not be valid near to
this critical field, which certainly coincides where the quantum effects
turns to be relevant. We are using the Lorentz-Dirac equation to introduce
the radiation reaction force and, as the Equivalence Principle is one of the
foundations of the Einstein General Relativity, no reference to Einstein
equations is made at any moment to develop this work, in order to avoid any
vicious causal recurrence.

PACS: 03.50.D
\end{abstract}

\section{Introduction\protect\smallskip\ }

There is a old classical problem that survived unsolved until now with a
quite general belief that it has no enough importance to be worthy to spend
time trying to understand it, sometimes claimed as unsolvable without a
quantal treatment$^{\text{\cite{Zuber},\cite{Souza}}}$. The problem in focus
is the effect of the radiation reaction force on the charged particle
dynamic, as given by the Lorentz-Dirac equation$^{\text{\cite{Dirac}-\cite
{Landau}}}$, which should affect the validity of the Equivalence Principle
at some circumstances$^{\text{\cite{Logunov}, \cite{Ginzburg}, \cite
{Rohrlich5}}}$.

Perhaps it is better to believe that any problem that survive such a long
time has to be important and it is to be worthwhile to spend any time as
necessary to figure out where we are wrong. In accordance to this belief,
there is now a renewed interest on this subject, with works pointing to
something new$^{\text{\cite{Rohrlich4}-\cite{Rohrlich6}}}$. By the way, we
have to notice that some different features are emerging from radiation
reaction force problem$^{\text{\cite{Denef}}}$. \qquad

There are a lot of works about this subject accumulated since the first
attempt was made by Dirac$^{\text{\cite{Dirac}}}$. However, the goal of this
work is not to discuss about these papers, but, instead, to add a
possibility to analyze this problem in a different perspective, with
emphasis in the Equivalence Principle, which validity is used as a good
starting point. From this point of view, one can show that what seems to be
uncomfortable, as the presence of radiation for charged particle performing
hyperbolic motion and its absence for one supported at rest in an uniform
gravitational field$^{\text{\cite{Fulton},\cite{Boulware}}}$, both
equivalent as the Equivalence Principle is valid, should lead to a new
physical feature performed by charged particles. As the subject of this work
is about the conditions of validity or not of the Equivalence Principle,
which is one of the foundations of the Einstein General Relativity, it is
just to notice that results from General Relativity are not used at any
moment, in order to avoid any possibility to fall in a vicious causal
recurrence.

What we have to do is to figure out the condition we have to provide such
that a charged particle can reach a stable hyperbolic motion. It is
necessary to furnish a subtle balance between an applied external force and
the radiation reaction force to get an hyperbolic motion regime. An
important consequence is that, taking account the Equivalence Principle, it
implies a gravitational mass that is slight greater than the inertial mass.
Perhaps because the main experimental justification that led Einstein to
formulate the Equivalence Principle, which is one of the foundations of his
General Theory of Relativity$^{\text{\cite{Steven}}}$, is the numerical
equality between inertial and gravitational mass, nowadays they are taken
quite as synonymous, so we have to be aware to avoid misleading conclusions.
It is very important to emphasize the central role played by the radiation
reaction force, so a slight upper deviation of gravitational mass should be
necessary to validate the Equivalence Principle.

However, a new feature emerges from an unexpected behavior of the
gravitational and inertial mass relation at a very strong gravitational
field, with the presence of a divergence that indicates a critical field
value. Such a strong gravitational field is just reachable at the early
stage of the universe or, perhaps, near massive black holes, and the
presence of the critical field value signs that the Equivalence Principle
should not be valid at a very strong gravitational field, and certainly it
coincides where the quantum effects turns to be relevant$^{\text{\cite
{Birrell}}}$.

Hyperbolic motion is the natural generalization of the concept of the
Newtonian uniformly accelerated motion due to a constant force applied to a
particle, which might be due to an uniform gravitational field. At
relativistic level, as the velocity is upper limited by the light velocity,
constant force don't imply constant acceleration; instead, it results the
above mentioned hyperbolic motion, which denomination comes from the
hyperbola that it is drawn in the $zt$-plane by this kind of motion.

An one dimensional hyperbolic motion of a particle of mass $m$ occurs as a
solution of the relativistic equation of motion$^{\text{\cite{Moller},\cite
{Landau}}}$ 
\begin{equation}
m\frac{d^{2}x^{\mu }}{d\tau ^{2}}=f^{\mu }(\tau )\ ,  \label{eq_motion}
\end{equation}
the relativistic force $f^{\mu }$ defined as 
\begin{equation}
f^{\mu }=\gamma \left( \frac{{\bf v\cdot F}}{c},{\bf F}\right) ,\ \gamma =%
\frac{1}{\sqrt{1-v^{2}/c^{2}}}\ ,
\end{equation}
when external force ${\bf F}$ is parallel to velocity ${\bf v}$ and it is
constant in the proper referential frame. Supposing the motion along the $z$%
-axes, the trajectory is given by 
\begin{equation}
(ct,\text{ }z)=\frac{c^{2}}{a}(\sinh \lambda \tau \text{, }\cosh \lambda
\tau )\ ,  \label{orbit}
\end{equation}
where $a=F/m$ is a constant proper acceleration and $\lambda =a/c$, $c$ the
velocity of light. Velocity and acceleration are given by 
\begin{equation}
(\stackrel{.}{z}^{0},\text{ }\stackrel{.}{z})=c(\cosh \lambda \tau \text{, }%
\sinh \lambda \tau )  \label{velocity}
\end{equation}
and 
\begin{equation}
(\stackrel{..}{z}^{0},\text{ }\stackrel{..}{z})=a(\sinh \lambda \tau \text{, 
}\cosh \lambda \tau )\ ,  \label{acceleration}
\end{equation}
respectively. The relativistic force responsible by the hyperbolic motion is 
\begin{equation}
f^{\mu }(\tau )=m(\stackrel{..}{z}^{0},\text{ }\stackrel{..}{z})=ma(\sinh
\lambda \tau \text{, }\cosh \lambda \tau )\ .  \label{force}
\end{equation}

The choice of the metric tensor $g^{\mu \nu }$ is such that $v^{\mu }v_{\mu
}=-c^{2}$ for four velocity $v^{\mu }=\stackrel{.}{x}^{\mu }$and, at non
relativistic limit, $a^{\mu }a_{\mu }=a^{2}$ for four acceleration $a^{\mu
}= $\/$\stackrel{..}{x}^{\mu }$.

\section{Hyperbolic motion of charged particles}

Equation of motion of a classical charged particle, including radiation
reaction force, is given by the well known Lorentz-Dirac equation 
\begin{equation}
ma^{\mu }(\tau )=f_{ext}^{\mu }(\tau )+m\tau _{0}\left( \stackrel{.}{a}^{\mu
}-\frac{1}{c^{2}}a^{\nu }a_{\nu }v^{\mu }\right) ,  \label{LD-eq}
\end{equation}
where $f_{ext}^{\mu }(\tau )$ is the external four-force and 
\begin{equation}
f_{rad}^{\mu }(\tau )=m\tau _{0}\left( \stackrel{.}{a}^{\mu }-\frac{1}{c^{2}}%
a^{\nu }a_{\nu }v^{\mu }\right) ,  \label{LD-force}
\end{equation}
where 
\begin{equation}
\tau _{0}=\frac{2}{3}\frac{e^{2}}{mc^{3}}\ ,  \label{tau-zero}
\end{equation}
is the Lorentz-Dirac relativistic radiation reaction force. The first term
is known as the Schott term$^{\text{\cite{Rohrlich}}}$ and it is responsible
by the well known non physical runaway solutions. The second is the Rohrlich
term, related to the power radiated 
\begin{equation}
{\cal R}=\frac{\text{ }dW_{rad}}{dt}=m\tau _{0}a^{\nu }a_{\nu }\ .
\label{rad-power}
\end{equation}

A well known condition for hyperbolic motion is 
\begin{equation}
\stackrel{.}{a}^{\mu }-\frac{1}{c^{2}}a^{\nu }a_{\nu }v^{\mu }=0\ ,
\label{hyperbole}
\end{equation}
which also implies $f_{rad}^{\mu }(\tau )=0$, so it seems to be easy to
produce hyperbolic motion of charged particles imposing a locally constant
external force, as in the uncharged particle case, but it could be a
mistake. To have a stable hyperbolic motion we have to get a very sensible
balance between external and radiation reaction force, and before it the
condition (\ref{hyperbole}) is not true. It means that what happen before is
very important to get a stable hyperbolic motion regime and, although we
have the same equation (\ref{eq_motion}) after that, the force $f^{\mu
}(\tau )$ is not just $f_{ext}^{\mu }(\tau )$ anymore. To figure out why,
let us consider the Lorentz-Dirac equation written as$^{\text{\cite{Jackson}}%
}$ 
\begin{equation}
m(1-\tau _{0}\frac{d}{d\tau })a^{\mu }=f_{ext}^{\mu }(\tau )-\frac{1}{c^{2}}%
{\cal R}v^{\mu }=K^{\mu }(\tau )\ .
\end{equation}

Formal expansion like 
\begin{equation}
(1-\tau _{0}\frac{d}{d\tau })^{-1}=1+\tau _{0}\frac{d}{d\tau }+\tau _{0}^{2}%
\frac{d^{2}}{d\tau ^{2}}+\cdots  \label{formal}
\end{equation}
enables us to get a formal solution of Lorentz-Dirac equation as 
\begin{equation}
ma^{\mu }(\tau )=\sum_{n=0}^{\infty }\tau _{0}^{n}\frac{d^{n}}{d\tau ^{n}}%
K^{\mu }(\tau )\ .  \label{solution}
\end{equation}
We can insert the mathematical identity 
\begin{equation}
\frac{1}{n!}\int_{0}^{\infty }s^{n}e^{-s}ds=1  \label{identity}
\end{equation}
to transform (\ref{solution}) in a second order integro-differential
equation 
\begin{equation}
ma^{\mu }(\tau )=\int_{0}^{\infty }e^{-s}K^{\mu }(\tau +\tau _{0}s)ds\ ,
\label{ID-eq}
\end{equation}
which shows a possible non causal behavior. In an explicit form, we have 
\begin{equation}
ma^{\mu }(\tau )=\int_{0}^{\infty }\left. (f_{ext}^{\mu }-\frac{1}{c^{2}}%
{\cal R}v^{\mu })\right| _{\tau +\tau _{0}s}e^{-s}ds\ .  \label{non-causal2}
\end{equation}

Actual hyperbolic motion implies (\ref{orbit}-\ref{force}), from which it is
easy to show that $f_{ext}^{\mu }\rightarrow f^{\mu }$ and, therefore, 
\begin{equation}
\int_{0}^{\infty }\left. (f^{\mu }-\frac{1}{c^{2}}{\cal R}v^{\mu })\right|
_{\tau +\tau _{0}s}e^{-s}ds=f^{\mu }(\tau )\ ,  \label{non-causal3}
\end{equation}
recovering the equation (\ref{eq_motion}), in accordance with equation (\ref
{LD-eq}) and condition (\ref{hyperbole}). But, while the motion is
approaching the hyperbolic regime, we have a limiting process 
\begin{equation}
\stackrel{.}{a}^{\mu }-\frac{1}{c^{2}}a^{\nu }a_{\nu }v^{\mu }\rightarrow
0\Rightarrow f_{rad}^{\mu }(\tau )\rightarrow 0\   \label{limit}
\end{equation}
such that the total force behaves to 
\begin{equation}
f_{ext}^{\mu }(\tau )+f_{rad}^{\mu }(\tau )\rightarrow f^{\mu }(\tau )\ .
\label{limit2}
\end{equation}
To have an hyperbolic motion, it is imperative that the applied external
force goes to 
\begin{equation}
f_{ext}^{\mu }(\tau )\rightarrow f_{ext}^{\mu }(\tau )=\int_{0}^{\infty
}f^{\mu }(\tau +\tau _{0}s)e^{-s}ds  \label{f-ext}
\end{equation}
as well as the radiation reaction force goes to 
\begin{equation}
f_{rad}^{\mu }(\tau )\rightarrow f_{Roh}^{\mu }(\tau )=\left.
\int_{0}^{\infty }\frac{1}{c^{2}}{\cal R}v^{\mu }\right| _{\tau +\tau
_{0}s}e^{-s}ds\   \label{f-Rohr}
\end{equation}
such that, after reaching hyperbolic motion, 
\begin{equation}
f^{\mu }(\tau )=f_{ext}^{\mu }(\tau )+f_{Roh}^{\mu }(\tau )\ ,
\label{balance}
\end{equation}
$f^{\mu }(\tau )$ given by (\ref{force}).

Using (\ref{force}), spatial component of equation (\ref{f-ext}) becomes 
\begin{equation}
f_{ext}(\tau )=\frac{ma}{2}\left( \frac{e^{\lambda \tau }}{(1-\lambda \tau
_{0})}+\frac{e^{-\lambda \tau }}{(1+\lambda \tau _{0})}\right)
\label{f-ext2}
\end{equation}
and, from equations (\ref{acceleration}) and (\ref{rad-power}), 
\begin{equation}
f_{Roh}(\tau )=-\frac{ma}{2}\lambda \tau _{0}\left( \frac{e^{\lambda \tau }}{%
(1-\lambda \tau _{0})}-\frac{e^{-\lambda \tau }}{(1+\lambda \tau _{0})}%
\right) \ .  \label{f-Rohr2}
\end{equation}

In the same way, time components become 
\begin{equation}
f_{ext}^{0}(\tau )=\frac{ma}{2}\left( \frac{e^{\lambda \tau }}{(1-\lambda
\tau _{0})}-\frac{e^{-\lambda \tau }}{(1+\lambda \tau _{0})}\right)
\label{p-ext2}
\end{equation}
and 
\begin{equation}
f_{Roh}^{0}(\tau )=-\frac{ma}{2}\lambda \tau _{0}\left( \frac{e^{\lambda
\tau }}{(1-\lambda \tau _{0})}+\frac{e^{-\lambda \tau }}{(1+\lambda \tau
_{0})}\right) .  \label{p-Rohr2}
\end{equation}
From equations (\ref{f-ext2}-\ref{p-Rohr2}), it is easy to see that the
total force (\ref{balance}) satisfies (\ref{force}) necessary to have an
hyperbolic motion.

It shows that the external force necessary to produce an hyperbolic motion
of neutral particle, 
\begin{equation}
f_{ext}(\tau )=f(\tau )=ma\cosh \lambda \tau \ ,  \label{neutral}
\end{equation}
is smaller than the external force (\ref{f-ext2}) necessary to give the same
hyperbolic motion of charged particle. All external force applied to neutral
particle is used to increase its kinetic energy, 
\begin{equation}
\frac{\text{ }dW}{dt}=vF=mav=mac\sinh \lambda \tau =mc^{2}\frac{\text{ }%
d\gamma }{dt}\ .  \label{pot}
\end{equation}
On the other hand, for charged particle, external force (\ref{f-ext2}), that
can be written as 
\begin{equation}
f_{ext}(\tau )=f(\tau )-f_{Roh}(\tau )\ ,
\end{equation}
provides the increase of kinetic energy in the same amount as given in (\ref
{pot}) and supplies, through $f_{Roh}(\tau )$, the energy lost carried by
radiation.

\section{Uniform gravitational field}

We saw in the previous section that charged particle performing hyperbolic
motion have to be submitted to an external force given by 
\begin{equation}
F_{ext}=\frac{ma}{2}\left( \frac{(1+\beta )}{(1-\lambda \tau _{0})}+\frac{%
(1-\beta )}{(1+\lambda \tau _{0})}\right) ,  \label{beta-force}
\end{equation}
where $F_{ext}(\tau )$ is the measurable force related to the relativistic
force by $f_{ext}(\tau )=\gamma F_{ext}(\tau )$.

The Equivalence Principle$^{\text{\cite{Rohrlich2},\cite{Rohrlich3}}}$ says
that a particle at rest in the laboratory frame $R_{lab}$ immersed in an
uniform gravitational field $g$ is seen by observer in a free falling
inertial frame $R$ as performing hyperbolic motion with local constant
acceleration $a=g$. The local constant force responsible by its hyperbolic
motion is the normal force $F_{n}=-F_{g}$ that supports the particle against
the gravitational force $F_{g}$, so, in absolute value it is equal to $mg$
for uncharged particle. But, for charged particle, the normal force $F_{n}$
must be equal to the external force $F_{ext}$ of equation (\ref{beta-force})
for $\beta =0$, 
\begin{equation}
F_{ext}\rightarrow F_{n}=\frac{mg}{(1-\lambda ^{2}\tau _{0}^{2})}\ .
\label{f-normal}
\end{equation}

This result suggests that the observer in the laboratory frame $R_{lab}$
measures the gravitational force acting on charged particle as 
\begin{equation}
F_{g}=\frac{mg}{(1-\lambda ^{2}\tau _{0}^{2})}\   \label{f-grav}
\end{equation}
such that

\begin{equation}
m^{\ast }=\frac{m}{(1-\lambda ^{2}\tau _{0}^{2})}\   \label{grav-mass}
\end{equation}
should define the gravitational mass $m^{\ast }$ of charged particle with
inertial mass $m$. For a typical charged particles as electron or proton, $%
\tau _{0}\simeq 0.62\times 10^{-23}s$ and $\tau _{0}\simeq 0.34\times
10^{-26}s$, respectively. So, in a field magnitude typical for a terrestrial
gravitational field, $g\simeq 10m/s$ and $\lambda =g/c\simeq 3.3\times
10^{-8}s^{-1}$, we can see that $\lambda ^{2}\tau _{0}^{2}\simeq $ $%
4.\,3\times 10^{-62}$ and $\lambda ^{2}\tau _{0}^{2}\simeq 2.2\times 10^{-69}
$, respectively for electron and proton, a very small numbers, such that $%
1-\lambda ^{2}\tau _{0}^{2}\cong 1$. As a consequence, the gravitational
mass is just slight greater than the inertial mass, $m^{\ast }\gtrsim m$,
and the gravitational and inertial mass relation defined by equation (\ref
{grav-mass}) is much as close to unit, $r_{g}=m^{\ast }/m\cong 1$. It means
that there is no consequence, for practical purpose, due to this slight up
deviation of gravitational mass in relation to the inertial mass, at least
in a region with gravitational field of magnitude as considered above. In
fact, it is true for a very long field interval, starting with $g=0$ and
going until to reach a very strong gravitational field of the order $%
g\thicksim 10^{30}m/s^{2}$, as we are going to show in the next step.

\begin{center}
figure

Figure caption: Gravitational and inertial mass relation, $r$, as function
of the gravitational field $g$. There is a critical point defined by $%
g_{c}=c/\tau _{0}$, which is particle dependent and has the value $%
g_{c}\simeq 8.8\times 10^{36}m/s^{2}$ for proton.
\end{center}

In a such strong gravitational field region, the field dependence of $r_{g}$%
, see equation (\ref{grav-mass}), starts to manifest, as we can see in the
figure 1. It increases very slowly and remains very close to unit, starting
with $g=0$ until to reach the very strong field magnitude of the order $%
g\thicksim 10^{30}m/s^{2}$, approaching the divergence point given by the
condition $1-\lambda ^{2}\tau _{0}^{2}=0$. Then, $r_{g}$ increases fast to
the infinite as the gravitational field goes to its critical value $%
g_{c}=c/\tau _{0}$. This critical field value is mass dependent, with $%
g_{c}\simeq 1.8\times 10^{35}m/s^{2}$ and $g_{c}\simeq 8.8\times
10^{36}m/s^{2}$ for electron and proton, respectively. Where is possible to
find such strong gravitational field? Certainly at the early time of the
Universe and, perhaps, near massive black holes hosted at the center of some
galaxies. Using the Newtonian formula of the gravitational field, $%
g=MG/R^{2} $, with $M\sim 10^{11}$ times the solar mass, $\sim 1.99\times
10^{30}kg$. A charged particle must approach as close as $R\approx 1.2\times
0^{-3}m\approx 1mm$ to see a gravitational field strong as $g_{c}$. The
matter density to give such strong field is of order $10^{30}kg/m^{3}$. The
divergence of the gravitational and inertial mass relation at the critical
field value signs that the Equivalence Principle should not be valid for
very strong gravitational field, and the quantum effects turns to be
relevant. It lasts some crude questions: what about general relativity at
this very strong gravitational field? or is it meaningful the concept of
unification of general relativity and quantum theory$^{\text{\cite{Birrell}- 
\cite{Penrose}}}$?

Above the critical point $g_{c}$, the relation $r_{g}$ turns to be negative,
with a possible physical interpretation to be investigated. Perhaps, it
should bring us some light to figure out a new physics required to explain
the recent data concerned to the cosmic rays with a very high energy, above $%
10^{19}eV$. There is a possibility that the primary particles of the ultra
energetic cosmic rays should be a relic of the charged particles scattered
in the early stage of the Universe$^{\text{\cite{Nagano}}}$.

\section{Conclusion}

Although the final equation of motion of neutral and charged particles in
hyperbolic motion regime seems to be identical, we realize that there is a
fundamental difference between them. For charged particle, actually, the
total locally constant force is the sum of an applied and the radiation
reaction forces in a subtle combination in such a way that results the same
as for neutral particle. An extra force is necessary to supply the same
kinetic energy as for uncharged particle plus the energy lost carried by
radiation. An interesting implication, as the Equivalence Principle is taken
account, is that the gravitational mass of charged particle must be greater
than its inertial mass in a very small amount. It is small enough to not be
detected by any experimental or practical devices, but it helps us to figure
out the condition necessary to get an hyperbolic motion regime and to
understand the meaning of its equivalence with a charged particle supported
at rest in an uniform gravitational field.

Until now the equality between gravitational and inertial mass was
understood as the essence of the Equivalence Principle, condition we
realized not to be true for charged particle, where, due to the presence of
radiation reaction, a slight deviation of gravitational mass compared to
inertial one is necessary to hold the Equivalence Principle. However, a new
feature comes from the gravitational and inertial mass relation behavior for
a very strong gravitational field, just reachable at the early stage of the
universe or, perhaps, near massive black holes. There exists a critical,
particle dependent, field value that signs a limit of validity of the
Equivalence Principle. It certainly coincides where the quantum effects
turns to be dominant.

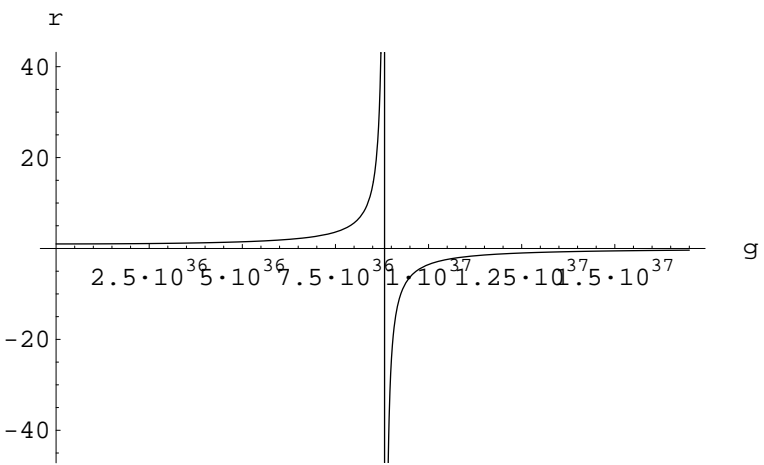
\end{document}